\documentstyle[12pt,aaspp4]{article}
\begin{document}
\newcommand{\up}[1]{\ifmmode^{\rm #1}\else$^{\rm #1}$\fi}
\newcommand{\zdot}{\makebox[0pt][l]{.}}
\newcommand{\upd}{\up{d}}
\newcommand{\uph}{\up{h}}
\newcommand{\upm}{\up{m}}
\newcommand{\ups}{\up{s}}
\newcommand{\arcd}{\ifmmode^{\circ}\else$^{\circ}$\fi}
\newcommand{\arcm}{\ifmmode{'}\else$'$\fi}
\newcommand{\arcs}{\ifmmode{''}\else$''$\fi}

\title{The Araucaria Project. The Distance to the Sculptor Galaxy NGC 247
from Near-Infrared Photometry of Cepheid Variables
\footnote{Based on observations obtained with the ESO VLT for 
Large Programme 171.D-0004}
}
\author{Wolfgang Gieren}
\affil{Universidad de Concepci{\'o}n, Departamento de Astronomia,
Casilla 160-C, Concepci{\'o}n, Chile}
\authoremail{wgieren@astro-udec.cl}
\author{Grzegorz Pietrzy{\'n}ski}
\affil{Universidad de Concepci{\'o}n, Departamento de Astronomia,
Casilla 160-C,
Concepci{\'o}n, Chile}
\affil{Warsaw University Observatory, Al. Ujazdowskie 4, 00-478, Warsaw,
Poland}
\authoremail{pietrzyn@hubble.cfm.udec.cl}
\author{Igor Soszy{\'n}ski}
\affil{Warsaw University Observatory, Al. Ujazdowskie 4, 00-478, Warsaw,
Poland}
\authoremail{soszynsk@astrouw.edu.pl}
\author{Olaf Szewczyk}
\affil{Universidad de Concepi{\'o}n, Departamento de Astronomia, Casilla 160-C,
Concepci{\'o}n, Chile}
\authoremail{szewczyk@astro-udec.cl}
\author{Fabio Bresolin}
\affil{Institute for Astronomy, University of Hawaii at Manoa, 2680 Woodlawn 
Drive, 
Honolulu HI 96822, USA}
\authoremail{bresolin@ifa.hawaii.edu}
\author{Rolf-Peter Kudritzki}
\affil{Institute for Astronomy, University of Hawaii at Manoa, 2680 Woodlawn 
Drive,
Honolulu HI 96822, USA}
\authoremail{kud@ifa.hawaii.edu}
\author{Miguel A. Urbaneja}
\affil{Institute for Astronomy, University of Hawaii at Manoa, 2680 Woodlawn
Drive, 
Honolulu HI 96822, USA}
\authoremail{urbaneja@ifa.hawaii.edu}
\author{Jesper Storm}
\affil{Astrophysikalisches Institut Potsdam, An der Sternwarte 16, D-14482
Potsdam, Germany}
\authoremail{jstorm@aip.de}
\author{Dante Minniti}
\affil{Departamento de Astronomia y Astrofisica, Pontificia Universidad Cat{\'o}lica
de Chile, Casilla 306, Santiago 22, Chile}
\authoremail{dante@astro.puc.cl}
\author{Alejandro Garc{\'i}a-Varela}
\affil{Universidad de los Andes, Departamento de Fisica, Bogot{\'a}, Colombia}

\begin{abstract}
We have obtained deep near-infrared images in J and K filters of four fields
in the Sculptor Group spiral galaxy NGC 247 with the ESO VLT and ISAAC camera.
For a sample of ten Cepheids in these fields, previously discovered by Garc{\'i}a-Varela
et al. from optical wide-field images, we have determined mean J and K magnitudes
and have constructed the period-luminosity (PL) relations
in these bands. Using the near-infrared PL relations together with those in
the optical V and I bands, we have determined a true distance modulus for NGC 247 of
27.64 mag, with a random uncertainty of $\pm$2\% and a systematic uncertainty
of $\sim$4\% which is dominated by the effect of unresolved stars on the Cepheid
photometry. The mean reddening affecting the NGC 247 Cepheids of E(B-V) = 0.18 $\pm$ 0.02 mag
is mostly produced in the host galaxy itself and is significantly higher than what 
was found in the previous optical Cepheid studies in NGC 247 of our own group,
and Madore et al., leading to a 7\% decrease in the previous optical Cepheid distance.
As in other studies of our project, the distance
modulus of NGC 247 we report is tied to an assumed LMC distance modulus of 18.50.
Comparison with other distance measurements to NGC 247 shows that the present
IR-based Cepheid distance is the most accurate among these determinations.

With a distance of 3.4 Mpc, NGC 247 is about 1.5 Mpc more distant than NGC 55
and NGC 300, two other Sculptor Group spirals analyzed before with the same
technique by our group.
\end{abstract}


\section{Introduction}

In our ongoing Araucaria project (Gieren et al. 2005a) we are improving
various stellar standard candles as tools for precise distance measurements.
The overall goal is a very accurate calibration of the distance ladder in the 
local Universe, out to perhaps 10 Mpc, which we hope to achieve by
tracing down the age and metallicity dependences of the various standard
candles we are investigating by comparative analyses of the distances
to the same targets we obtain from the different methods. This will lay
the foundation for a truly accurate determination of the Hubble constant,
reducing the uncertainty on Ho achieved by the HST Key Project on the
Extragalactic Distance Scale (Freedman et al. 2001).
 The methods we use include Cepheid variables,
via Period-Luminosity (PL) relations in the optical and infrared spectral 
domains (e.g. Pietrzynski et al. 2002; Gieren et al. 2005b), the K-band Period-Luminosity -Metallicity 
relation for RR Lyrae
stars (e.g. Pietrzynski et al. 2008; Szewczyk et al. 2008), near-infrared magnitudes of red clump
giants (e.g. Pietrzynski and Gieren 2002; Pietrzynski, Gieren and Udalski 2003), 
the Flux-Weighted
Gravity-Luminosity Relation (FGLR) for blue supergiant stars (e.g. Kudritzki et al.
2008), and most recently late-type eclipsing binary systems in the LMC
(Pietrzynski et al. 2009) which bear the potential to determine the LMC
distance, as a crucial step in the distance ladder, with an accuracy
of 1 \%. 

In this paper, we are using near-infrared observations of Cepheids
in the spiral galaxy NGC 247 located in the nearby Sculptor Group,
 to determine an accurate distance
to this galaxy which improves on the preliminary distance we had derived
from optical (VI) data in a previous paper (Garc{\'i}a-Varela et al. 2008; hereafter 
Paper I).
Infrared photometry has a number of distinct advantages over optical photometry
for distance work with Cepheid variables. The most important gain is the very
substantial reduction of the sensitivity of the observed Cepheid magnitudes to
interstellar extinction. In addition, the intrinsic dispersion of the Cepheid
PL relation becomes smaller towards the infrared region of the spectrum
as compared to optical passbands, helping
in obtaining more accurate distances. A further important advantage
of the near-infrared domain is the fact that the mean magnitudes of Cepheid
variables can be obtained with an accuracy of about 2 \% from just one observation
obtained at an arbitrary pulsation phase with the technique developed by
Soszynski et al. (2005). This technique can be applied when an optical light curve
(V and/or I) of a Cepheid, and its period, are known from a previous discovery
survey, which has to be conducted in the optical spectral range in which Cepheid
amplitudes are significantly larger than in the infrared, making their discovery
much easier. In the Araucaria Project it
has therefore been our strategy to discover the Cepheid populations in our
Local and Sculptor Group target galaxies from wide-field imaging surveys in
optical (BVI) bands, and select sub-samples of the detected Cepheid for
follow-up observations in the infrared. We have shown in previous papers
that a combination of the distance moduli derived in the optical VI
and near-infrared JK bands leads to a very accurate determination of the
true, absorption-corrected distance modulus of the host galaxy, 
and the mean color excess affecting
its Cepheids (Gieren et al. 2008a,b; Soszynski et al. 2006; Gieren et al. 2006;
Pietrzynski et al. 2006a; Gieren et al. 2005b).
Generally, we have found in these previous studies
that the intrinsic reddening found in the spiral, and even the smaller irregular
Local Group galaxies we have studied is larger than previously assumed
by other authors, which has demonstrated that infrared work with its potential
to virtually eliminate reddening as a significant source of systematic error
is indeed imperative
if one seeks to achieve distance accuracies in the 3-5\% range, which is
the aim of our project.

The principal properties of NGC 247 have been described in Paper I. At a distance of
about 4 Mpc it is possible, using 8m-class telescopes, to 
obtain high-quality infrared photometry for the long-period Cepheid population
we have detected in this galaxy (Paper I), and high-quality
low-resolution spectra for its blue supergiant population as needed for the
FGLR analysis (Kudritzki et al. 2008; Urbaneja et al. 2008). The Cepheid and
blue supergiant methods of distance measurement can therefore be well compared
on this galaxy. 

Our paper is organized as follows. In section 2 we describe the near-infrared
observations of the Cepheids in NGC 247, the data reductions, and the
calibration of our photometry. In section 3 we derive the J- and K-band Cepheid 
PL relations
in NGC 247 obtained from our data and determine the true distance modulus 
to the galaxy from a multiwavelength analysis. In section 4 we discuss our
results, and in section 5 we summarize our conclusions.

\section{Observations, Data Reduction and Calibration}

We used deep J- and K-band images recorded with the 8.2 m ESO VLT equipped with
the Infrared Spectrometer and Array Camera (ISAAC). Figure 1 shows the location
of the four 2.5 x 2.5 arcmin fields observed in service mode during 7 nights
between August 9  and November 18,  2005. The location of the fields was chosen in such a way
as to maximize the number of Cepheids observed and optimize their period
distribution. Fields 1, 2 and 4, were observed in both NIR bands two times, on two
different nights, and therefore at different pulsation phases of the Cepheids in
these fields. Field 3 was observed on a single night only.

The observations were carried out using a dithering technique,
with a dithering of the frames following a random pattern characterized by
typical offsets of 15 arcsec. In order to perform an accurate sky
subtraction for our frames, we frequently observed random comparison
fields located outside the main body of the galaxy, where the stellar
density was very low, using the AutoJitterOffset template.
The final frames in the J and K bands were obtained
as a co-addition of 34 and 238 single exposures obtained with integration times 
of 30 and 15 s, respectively. Thus, the total exposure time for a given observation
was 17 minutes in J and 57 minutes in K. The observations were obtained under 
excellent seeing conditions, typically around 0.5 arcsec. Standard stars on the
UKIRT system (Hawarden et al. 2001) were observed along with the science exposures
to allow an accurate transformation of the instrumental magnitudes to the
standard system.

The images were reduced using the program JITTER from the ECLIPSE package developed
by ESO to reduce near-IR data. The point-spread function (PSF) photometry was
carried out with the DAOPHOT and ALLSTAR programs. The PSF model was derived 
iteratively from 20 to 30 isolated bright stars following the procedure described
by Pietrzynski, Gieren \& Udalski (2002). In order to convert our profile photometry to the
aperture system, aperture corrections were computed using the same stars as those
used for the calculation of the PSF model. The median of the aperture corrections
obtained for all these stars was finally adopted as the aperture correction for
a given frame. The aperture photometry for the standard stars was performed with
DAOPHOT using the same aperture as the one adopted for the calculation of the
aperture corrections. 

The astrometric solution for the observed fields was performed by cross-identification 
of the brightest stars in each field with the Infrared Digitized Sky Survey 2 
(DSS2-infrared) images. We used programs developed by Udalski et al. (1998) to
calculate the transformations between the pixel grid of our images and equatorial
coordinates of the DSS astrometric system. The internal error of the transformation is 
less than 0.3 arcsec, but systematic errors of the DSS coordinates can be up to
about 0.7 arcsec. 

In order to perform an external check on our photometric zero points, we tried
to compare the magnitudes of stars in the Two Micron All Sky Survey (2MASS) Point
Source Catalog located in our NGC 247 fields with our own photometry. Unfortunately,
even the brightest stars in our data set whose photometry is not affected by
nonlinearity problems are still very close to the faint magnitude limit of the
2MASS catalog and have 2MASS magnitudes with formal errors of $\sim$0.2 mag. 
Moreover, all our NGC 247 fields are located in regions of high stellar density 
(see Fig. 1), and most of the 2MASS stars turn out to be severely blended, as seen
at the higher resolution of our VLT ISAAC images. It was therefore not possible
to carry out a reliable comparison of the two photometries. However, since our
reduction and calibration procedure is extremely stable, and since the set of UKIRT
standard stars observed is identical to the one used in our previous studies of
NGC 300 (Gieren et al. 2005b) and NGC 55 (Gieren et al. 2008a),
we should have achieved
the same typical zero point accuracy we were able to achieve in our previous
NIR studies of Local Group galaxies where a comparison with 2MASS magnitudes for
a common set of stars was possible. From this argument, we expect that our
photometric zero points are determined to better than $\pm$0.03 mag in both J
and K filters. 

Our observed fields in NGC 247 contain a subset of ten of the 23 Cepheids we have
previously discovered in NGC 247 (see Paper I). All individual observations in J 
and K filters we obtained for these variables are presented in Table 1 which
lists the star's IDs, heliocentric Julian day of the observations, and the
measured magnitudes in J and K with their respective standard deviations (as returned
by DAOPHOT). For most
of the Cepheids we collected two observations per filter. For four objects we 
obtained only one JK observation because of their location in field 3.

\section{Near-infrared period-luminosity relations and distance determination}

All individual J and K measurements reported in Table 1 were transformed to the
mean magnitudes of the Cepheids using the recipe given by Soszynski et al. (2005). 
The corrections to derive the mean magnitudes from the observed random-phase 
magnitudes were calculated by taking advantage of the V-and I-band light curves
of the variables from Paper I, exactly in the same way as described in the
Soszynski et al. paper. For all Cepheids in the sample, the mean J and K
magnitudes derived from the two individual observations at different pulsation phases 
agree very well. We were helped by the fact that the optical photometric observations
in Paper I and the NIR follow-up observations reported in this paper were obtained
relatively close in time (1-2 years), reducing the effect any inaccuracy in the
periods of the Cepheids will have on the calculations of the phases of the more
recent NIR data. For the Cepheids with two sets of JK observations obtained on
different nights, the difference between the mean magnitudes
calculated from the two independent random-phase JK observations was comparable
to the standard deviations of the individual observations. 

Table 2 gives the final intensity mean J and K magnitudes of the NGC 247
Cepheids we observed. The periods in the Table were taken from Paper I. The
uncertainties on the mean magnitudes in Table 2 contain a 0.03 mag contribution
coming from the transformation of the random phase to the mean magnitudes, as
found appropriate in the study of Soszynski et al. (2005). In Fig. 2, we plot
the positions of the Cepheids in Table 2 in the K, J-K Color-magnitude diagram
for the stars in NGC 247. It is seen that the variables populate the expected
region in the CMD and clearly delineate the Cepheid instability strip.

In Figures 3 and 4, we show the J-and K-band period-mean magnitude diagrams 
obtained from the data in Table 2. In both diagrams, the Cepheids define
very tight PL relations. Least-squares fits to a line yield slopes of 
-2.96 $\pm$ 0.19 in the J band, and -3.19 $\pm$ 0.22 in the K band. Due to
the relatively low number of Cepheids in these diagrams, the errors on the
slopes are large, but the slope values agree within the uncertainties with the 
corresponding slopes on these diagrams determined for the LMC by Persson et al. (2004), 
which are (-3.153 $\pm$ 0.051) in J, and (-3.261 $\pm$ 0.042) in K, respectively. 
We are therefore justified to apply
the same procedure we have used in all our previous Cepheid IR studies, which is
to adopt the extremely well-determined LMC PL relation slopes of Persson et al.
in our fits to find the zero points from our data. Doing this, we obtain the
following J- and K-band Cepheid PL relations in NGC 247: \\

J = -3.153 log P + (25.635 $\pm$ 0.039)  \hspace*{1cm}   $\sigma$ = 0.12 \\

K = -3.261 log P + (25.238 $\pm$ 0.040)  \hspace*{1cm}   $\sigma$ = 0.13 \\

In order to determine the relative distance moduli between NGC 247 and the LMC,
we need to convert the NICMOS (LCO) photometric system used by Persson et al. (2004)
to the UKIRT system used in this paper. According to Hawarden et al. (2001), there
are just zero point offsets between the UKIRT and the NICMOS systems (e.g. no
color dependences) in the J and K filters which amount to (0.034 $\pm$ 0.004) mag, and 
(0.015 $\pm$ 0.007) mag, respectively. 
Applying these offsets, and assuming a true distance modulus of 18.50 for the LMC
as in our previous work in the Araucaria Project, we derive distance moduli for
NGC 247 of 27.799 $\pm$ 0.038 mag in the J band, and 27.702 $\pm$ 0.04 mag
in the K band. The quoted uncertainties are those from the fits and do not include
systematic uncertainties, which will be discussed in the next section.

As in our previous papers in this series, we adopt the extinction law of Schlegel
et al. (1998) to find $R_{\lambda}$ (values given in Table 3) and fit a straight line to the relation
$(m-M)_{0} = (m-M)_{\lambda} - A_{\lambda} = (m-M)_{\lambda} - E(B-V) * R_{\lambda}$.
Using the distance moduli for NGC 247 in the V and I photometric bands derived
in Paper I, which are also given in Table 3, together with the values for the J and K bands 
derived above, we obtain from the least squares fit shown in Fig. 5 the following values
for the reddening, and the absorption-corrected, true distance modulus of NGC 247: \\

$ E(B-V) = 0.177 \pm 0.020$

$ (m-M)_{0} = 27.644 \pm 0.036 $

corresponding to a distance of NGC 247 of 3.38 $\pm$ 0.06 Mpc. The small uncertainties on both
the reddening, and the true distance modulus derived from the fit in Fig. 5 demonstrate
that these quantities are indeed very well determined by our data.
The agreement between the true distance moduli obtained from the
different bands, which are listed in Table 3, is excellent; in V, J and K the values differ less 
than 0.02 mag
from the adopted true distance modulus. Only in the I band the agreement is
slightly worse (0.05 mag), but clearly within the error bar of this determination.

\section{Discussion}

The Cepheid distance to NGC 247 derived in this paper is in principle affected
by a number of systematic uncertainties, which will now be discussed in turn.
The most serious concern with Cepheid-determined distances to late-type galaxies
is always with reddening. Classical Cepheids, as young stars, tend to be embedded in
dusty regions in their host galaxies, which makes a precise determination
of absorption corrections absolutely necessary. Since reddening is expected
to be patchy and individual Cepheids can thus be expected to have widely different
reddenings, the only way to reduce this problem is to include infrared photometry
in the distance analysis, as we are doing in our project. With the results
derived from the data in Table 3 we believe that any remaining uncertainty
on the true distance modulus due to reddening is smaller than 0.03 mag, in agreement
with the conclusions we had reached for the other target galaxies of the
Araucaria Project whose distances were determined from Cepheid photometry
in the VIJK bands (NGC 55, NGC 300, IC 1613, NGC 6822, WLM and NGC 3109; references
as given in the Introduction of this paper). It is worthwhile mentioning that
from Fig. 5 there is evidence that our assumed Galactic reddening law is valid,
to a very good approximation, in NGC 247 as well. This is in agreement with
our findings in the other spiral and irregular galaxies we have studied so far.

Very recently, Madore et al. (2009) have published a Cepheid distance to NGC 247
based on optical (VRI) CCD images obtained at the CTIO some 25 years ago. Their
distance result of 27.81 $\pm$ 0.10 mag is in near-perfect agreement with our
distance reported in Paper I based on the same filters (V and I). The mean reddening
they derive for the Cepheids in NGC 247 from their data of E(V-I)=0.07 $\pm$0.04,
corresponding to E(B-V)=0.12, is also almost identical to our own value of E(B-V)=0.13 
found in Paper I. Their sample of nine Cepheids
contains six objects independently discovered in our own survey, and the light curve
agreement is excellent for all common variables
with the exception of one object, cep020 (P=65.862 days) in our catalog in Paper I. 
The long time difference between the Madore et al. and our own observing
epochs indicates that both period, and the mean brightness of this peculiar Cepheid
have changed over the past $\sim$25 years. This peculiar Cepheid is not contained
in our present infrared-observed fields, and therefore not a source of systematic
uncertainty in our present distance moduli derived in the J and K bands. 

The excellent agreement of the results of Madore et al. (2009) with the results
obtained in Paper I is reassuring and constitutes a valuable external check of our own
previous results from optical photometry. The inclusion of infrared photometry in
this paper shows again (see previous papers in this series) that this is an absolutely
essential step to determine the total Cepheid reddening with a truly high accuracy. 
The decrease of the true distance modulus of NGC 247 found in this paper as compared to 
Paper I, and the Madore et al. (2009) result, is just a consequence of an
underestimate of the mean Cepheid reddening in both papers based on optical data alone
which provide a wavelength base which is just too small, even with very good data,
to assess the reddening with reasonable accuracy. This is also seen in Fig. 5; the
slightly outlying distance modulus in I produced the smaller reddening, and larger
distance modulus we had obtained in Paper I from the V and I data only. 

As in the previous papers in our project, we have applied utmost care to
determine the zero points of our photometry as accurately as possible. From the
arguments given in section 2 and our previous experience we believe that the
zero points are accurate to better than $\pm$0.03 mag, in both J and K bands.
An issue of concern is the crowding of stars in the images of NGC 247 which
could cause significant blending for some of Cepheids, increasing their
observed fluxes. To minimize this problem, we chose our observed fields in NGC 247
in the least crowded regions of the galaxy (see Fig. 1).
Fortunately, our VLT images are of exquisite quality and the
seeing during the exposures did never exceed 0.6 arcsec (for most images it was
0.4-0.5 arcsec). For none of the ten Cepheids in Table 2 there is evidence
for a photometrically significant companion star in the infrared images. This
is also borne out in the PL diagrams shown in Figs. 3 and 4, where no obvious outlier
is observed. A significantly blended Cepheid should be too bright for its period
and stand out from the mean PL relation, and this is clearly not observed in
the case of our sample. The only candidate, from its location in the K-band
PL plane, would be the variable cep016 which is $\sim$0.25 mag brighter than
the ridge line at the corresponding period, but such an offset is compatible
with the combined effect of the intrinsic width of the Cepheid instability strip,
which is about $\pm$0.2 mag in the K band,
and the standard deviation of the mean K magnitude of this star of $\pm$0.07 mag
(see Table 2). Also, this star lies almost exactly on the ridge line in 
the J-band period-luminosity plane, arguing against significant blending.
 As a general guide for the effect of blending on Cepheid distances
we can use our previous HST-based study on NGC 300 which is also a member galaxy 
of the
Sculptor Group. Comparison of HST ACS photometry to ground-based photometry
allowed us to set an upper limit of $\pm$0.04 mag for the effect blending of
the Cepheids in NGC 300 might have on the derived Cepheid distance modulus
(Bresolin et al. 2005). NGC 300, at about 1.9 Mpc (Gieren et al. 2005), is
about 1.5 Mpc nearer than NGC 247 however, so we might expect a slightly larger effect
of blending for NGC 247. From the arguments given, we assume that 0.06 mag,
or 3\% is a reasonable upper limit for the possible remaining effect of unresolved
stars in the Cepheid photometry on the distance result. The effect acts to make
Cepheids too bright and therefore tends to decrease the derived distance.

Another potential cause of concern is the possible non-universality of the 
Cepheid PL relation. Sandage and Tammann (2008) have recently put forward arguments
to demonstrate that both slope and zero point of the PL relation (in any given
passband) depend on the slope of the Cepheid instability strip on the HR diagram,
which in turn is metallicity-dependent. They conclude that as a consequence
the coefficients of the PL relation must depend, to some degree, on the
metallicity of the Cepheids. In their paper, they show a table which reports
empirically determined PL relation slopes in different galaxies which indeed seem to
suggest that the Cepheid PL relation is steepest for the most metal-rich
galaxies, supporting their claim from theoretical arguments. However, the
uncertainties on the empirical slopes of the PL relations in the different galaxies
they compare are not given; from our own work in the Araucaria Project, which
has generally provided the best Cepheid samples in the different target galaxies 
of the project, we know that the uncertainties attached to the least-squares fits
to the period-magnitude diagrams are generally large enough to hide any
systematic change of slope with metallicity, should it exist. So the empirical
evidence for a systematic change of the slope of the PL relation (in any band)
with metallicity is very weak, at the present time. On the other hand, the
OGLE project has found, from many hundreds of extremely well-observed Cepheids
in both Magellanic Clouds, that the slopes of the Cepheid PL relations in 
V and I, respectively in both Clouds agree very well, within extremely small uncertainties
(Udalski et al. 1998; Udalski 2000),
in spite of the metallicity difference of $\sim$0.4 dex between its Cepheid
populations (Luck et al. 1998), arguing against a dependence of the PL relation
slope on metallicity. Also, very recent and improved work using the infrared
surface brightness technique (Fouqu{\'e} and Gieren 1997; Gieren et al. 1997)
to measure direct distances to Cepheids in the Milky Way galaxy and the LMC
has led to a re-calibration of the method which has produced PL relations
in the LMC and Milky Way, in passbands from B through K, whose respective slopes agree
within a fraction of their respective standard deviations 
(Gieren et al. 2005c; Fouqu{\'e} et al. 2007). It seems to us that the
accumulated empirical evidence favors a universal, metallicity-independent
slope over the scenario put forward by Sandage and Tammann, at the present time,
but further work on this crucial question is obviously required. Until truly
convincing empirical evidence becomes available demonstrating a significant metallicity
effect on the slope of the PL relation, we will assume the constancy of the
slope in all photometric bands, in agreement with the results we have obtained
so far in the Araucaria Project.

Regarding the effect metallicity has on the {\it zero point} of the Cepheid PL relation,
a  recent study of Romaniello et al. (2008) 
indicates that there is indeed a small effect in optical bands, but only a marginal
effect in the near-infrared K band. In V, metal-rich Cepheids appear to be slightly
intrinsically fainter than their more metal-poor counterparts. The smallness and
sign of the metallicity effect on the Cepheid PL relation zero points found by
Romaniello et al. is in agreement with most other published work (discussed in that paper).
While we will re-discuss the metallicity effect from our own data on the Araucaria
project galaxies in the near future, our preliminary findings on the metallicity
dependence of the PL relation are consistent with a very small, and possible
vanishing effect on the zero points in optical and near-IR bands.
Any systematic uncertainty on the current distance modulus of NGC 247 from the
metallicity difference between its Cepheids and those of the LMC which provide
the fiducial PL relation is expected to be very small, probably less than 2\%.
This statement rests on both the smallness of the metallicity effect as found by 
Romaniello et al. (2008) and other authors, and the expectation that the mean metallicity
of the young stellar population in NGC 247 is quite close to that of the LMC.
This second statement is in turn based on the similarity of NGC 247 to NGC 300,
whose young stars (and H II regions) at mean galactocentric radii have been found
to possess LMC metallicity (Kudritzki et al. 2008; Bresolin et al. 2009).

As in the previous papers in this series, the distance of our target galaxy 
in this paper, NGC 247, is tied to an {\it assumed} distance modulus of 18.50 mag
for the LMC. In spite of the intense work of many groups over the years, this
value may still be uncertain by as much as 10\% (e.g. Benedict et al. 2002; 
Schaefer 2008). Very recently, the discovery of a number of late-type eclipsing
binary systems in the LMC for which the distances can be measured very
accurately has opened a new possibility to measure the distance to the LMC
with an unprecedented accuracy (Pietrzynski et al. 2009). This new route
might finally allow to measure the LMC distance with an accuracy close to 1\%.

>From the discussion in this section we conclude that the total systematic
uncertainty on the new infrared-based Cepheid distance reported in this paper
is in the order of $\sim$4\%. In estimating this number, we assume that future
work will confirm the universality of the slope of the Cepheid PL relation.
 The 4\% systematic error of our current distance
determination does not include the current uncertainty of the adopted LMC distance.

Among the various attempts which were made in the past to determine the distance  
to NGC 247 (which have been listed and discussed in Paper I), the present
determination is clearly the most accurate one. Probably the most accurate
previous measurement with methods independent of Cepheids
was made by Karachentsev et al. (2006) with the TRGB
method; their distance result of 27.87 $\pm$ 0.21 mag agrees with our infrared
Cepheid distance within the combined 1 $\sigma$ error bars of both measurements.
The fact that it agrees even better with our previous Cepheid distance to NGC 247
from VI data is certainly fortituous and may have its explanation in the
systematic uncertainties affecting the TRGB method.

The distance of NGC 247 determined in this paper confirms that this galaxy is
quite significantly more distant than the other two member galaxies of the
Sculptor Group we have studied before, NGC 55 (Gieren et al. 2008a; Pietrzynski
et al. 2006b) and NGC 300 (Gieren et al. 2005b). This confirms the previous
conclusion of Jerjen et al. (1998) that the Sculptor Group exhibits a filament-like
structure with a large depth extension in the line of sight.

\section{Summary and conclusions}

We have carried out the first distance determination for the
spiral galaxy NGC 247 located in the nearby Sculptor Group using deep 
near-infrared images, and our multiwavelength technique presented and applied
in the earlier studies in this series. The distance we determine has a random
uncertainty less than 2\%, and a systematic uncertainty in the order of $\pm$4\%.
The systematic uncertainty is likely to be dominated by the effect of unresolved
stars on the Cepheid photometry while the effect of metallicity on the Cepheid PL
relation, and particularly the effect of interstellar absorption are only smaller
contributors to the total systematic uncertainty of our result. The full assessment
of the reddening suffered by the NGC 247 Cepheids due the use of near-infrared
photometry has led to a decrease of the distance modulus reported in Paper I,
and independently by Madore et al. (2009), by 7\%.
As in all our previous studies in the Araucaria Project, the distance to
NGC 247 is tied to an assumed distance modulus of 18.50 for the LMC.

We find a mean reddening of E(B-V)=0.18 mag for the Cepheids in NGC 247. Since the
foreground reddening to this galaxy is only $\sim$0.02 mag (Schlegel et al. 1998),
0.16 mag are produced inside NGC 247, which is higher than in any other galaxy
we have studied so far in the Araucaria Project. This shows that it is mandatory
to use deep infrared photometry to handle the absorption problems when using
Cepheids for distance determinations to relatively massive spiral galaxies.

The present study adds another spiral galaxy to our sample of galaxies in
the Araucaria Project for which an accurate Cepheid distance, based on deep
infrared photometry, is now available for comparative analyses with the
distances obtained from other stellar techniques, particularly the Tip of
the Red Giant Branch method (e.g. Rizzi et al. 2006), and the FGLR method
(Kudritzki et al. 2003, 2008). Such comparative analyses will improve the
determination of the environmental dependences of the different techniques
of distance measurement.

We find that NGC 247 is about 1.5 Mpc more distant than the Sculptor Group spirals
NGC 300 and NGC 55, and therefore not physically associated to neither of these
two spirals.

\acknowledgments
WG, GP and DM gratefully acknowledge financial support for this
work from the Chilean Center for Astrophysics FONDAP 15010003, and from
the BASAL Centro de Astrofisica y Tecnologias Afines (CATA) PFB-06/2007. 
Support from the Polish grant N203 002 31/046 and the FOCUS
subsidy of the Fundation for Polish Science (FNP) is also acknowledged.
 I.S. was supported by the Foundation for Polish Science through the Homing Programme.
It is a great pleasure to thank the support astronomers at ESO-Paranal
for their expert help in the observations, and the ESO OPC for the generous
amounts of observing time at the VLT allocated to our Large Programme.

\begin{figure}[p] 
\vspace*{18cm}
\includegraphics{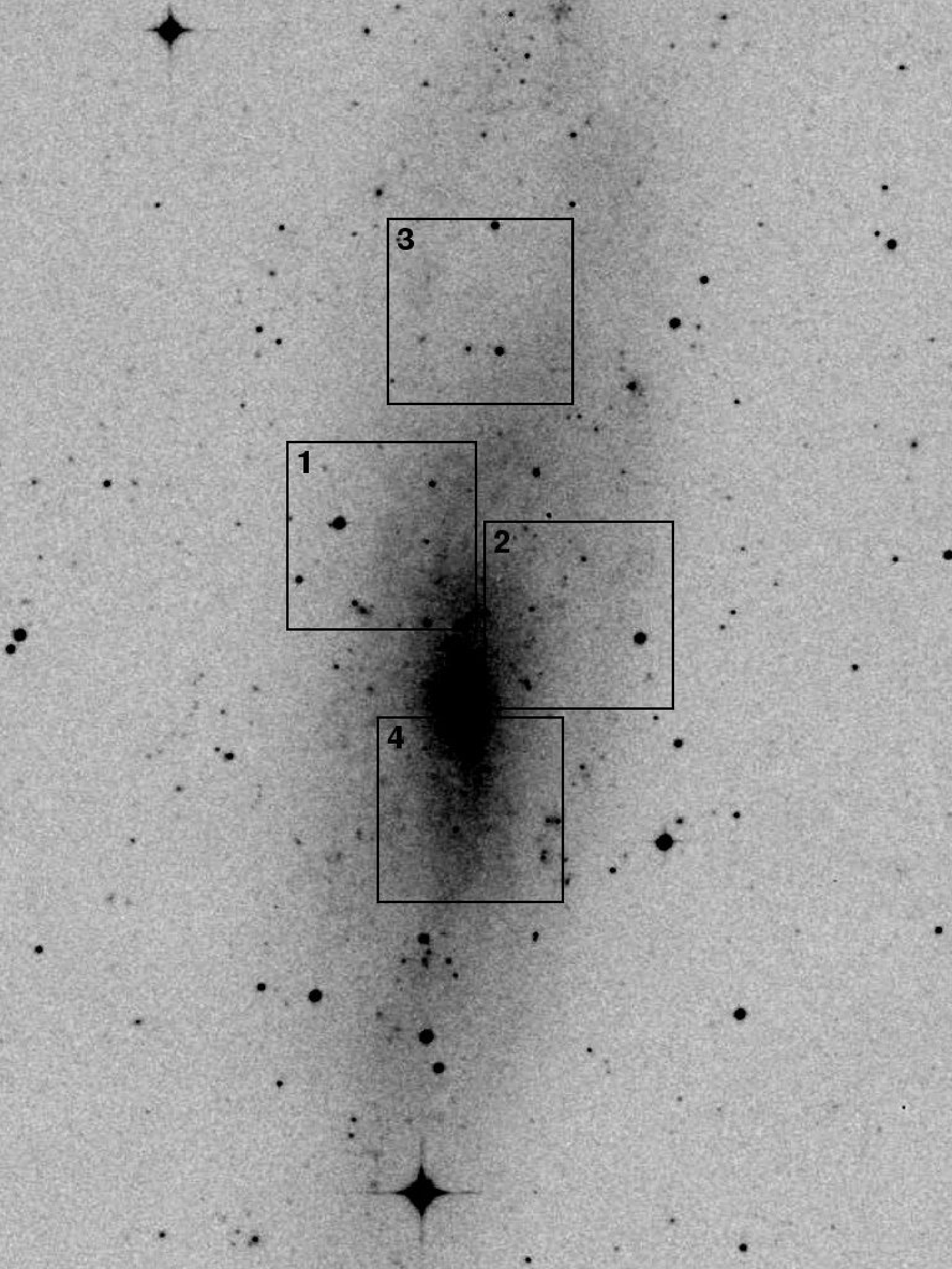} 
\caption{The location of the four observed VLT ISAAC fields in NGC 247 on the DSS
blue plate. These fields contain 10 Cepheids previously discovered in our
optical survey for Cepheid variables in this galaxy (Paper I). 
Each field except field No.3 was observed
twice on different nights to obtain the Cepheid infrared magnitudes
at two different pulsation phases.}
\end{figure}  

\begin{figure}[htb]
\vspace*{10cm}
\includegraphics{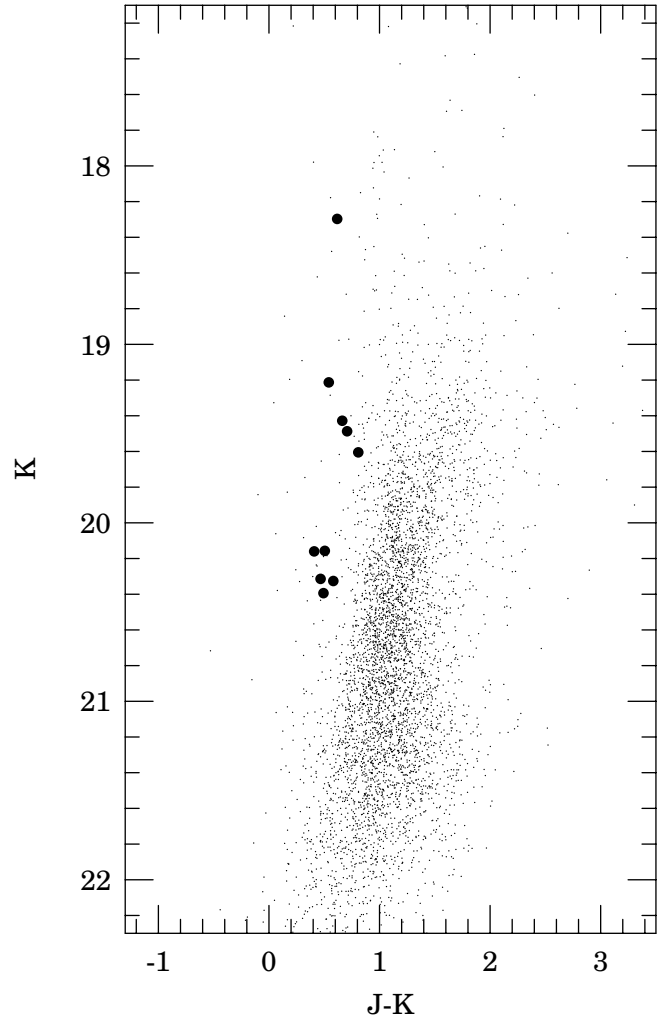}
\caption{The near-infrared color-magnitude diagram for NGC 247 obtained from our
VLT-ISAAC data. Also plotted are the positions of the Cepheids in our sample (filled circles). 
They delineate the location of the Cepheid instability strip in its
expected position. There is no evidence for any peculiarity of the Cepheids
from their locations in this diagram. }
\end{figure}

\begin{figure}[htb]
\vspace*{15cm}
\includegraphics{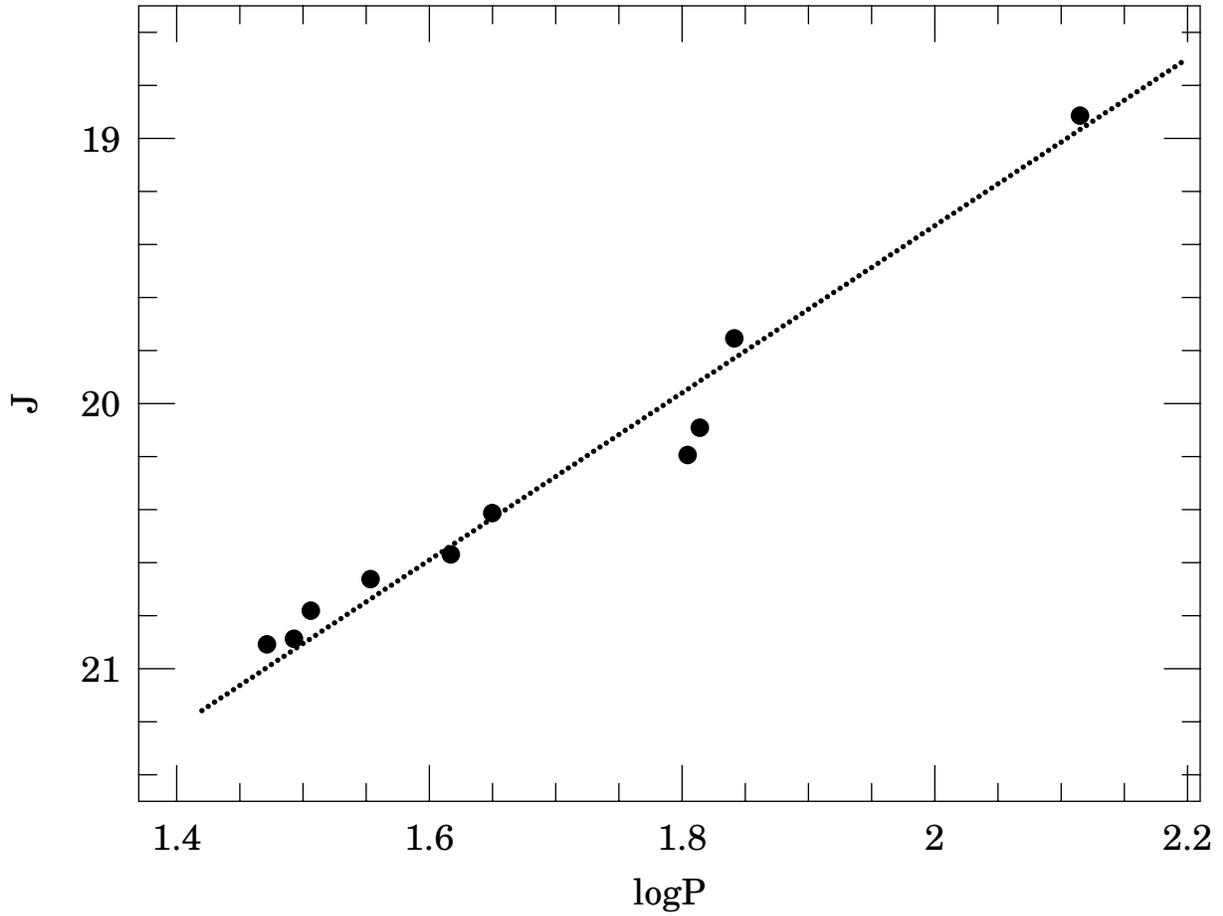}
\caption{The Cepheid period-luminosity relation for NGC 247 in the J  band,
as obtained from the mean magnitudes in Table 2. The dotted line is the
best fit to the data, with the slope fixed to the LMC value of Persson et al.
The mean magnitudes were determined using the technique of Soszynski et al.}
\end{figure}

\begin{figure}[htb]
\vspace*{15cm}
\includegraphics{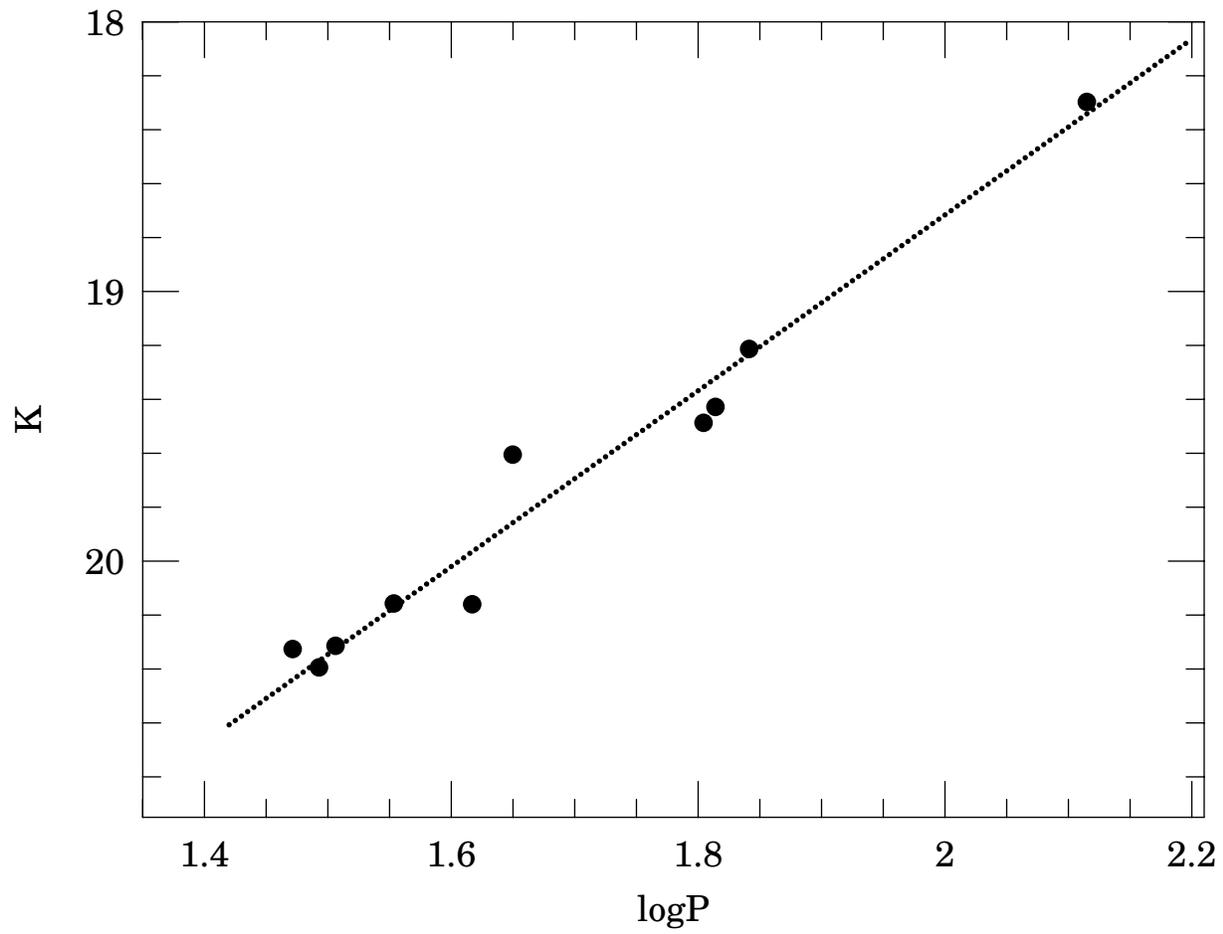}
\caption{Same as Fig. 3, for the K band.}
\end{figure}

\begin{figure}[p]
\includegraphics{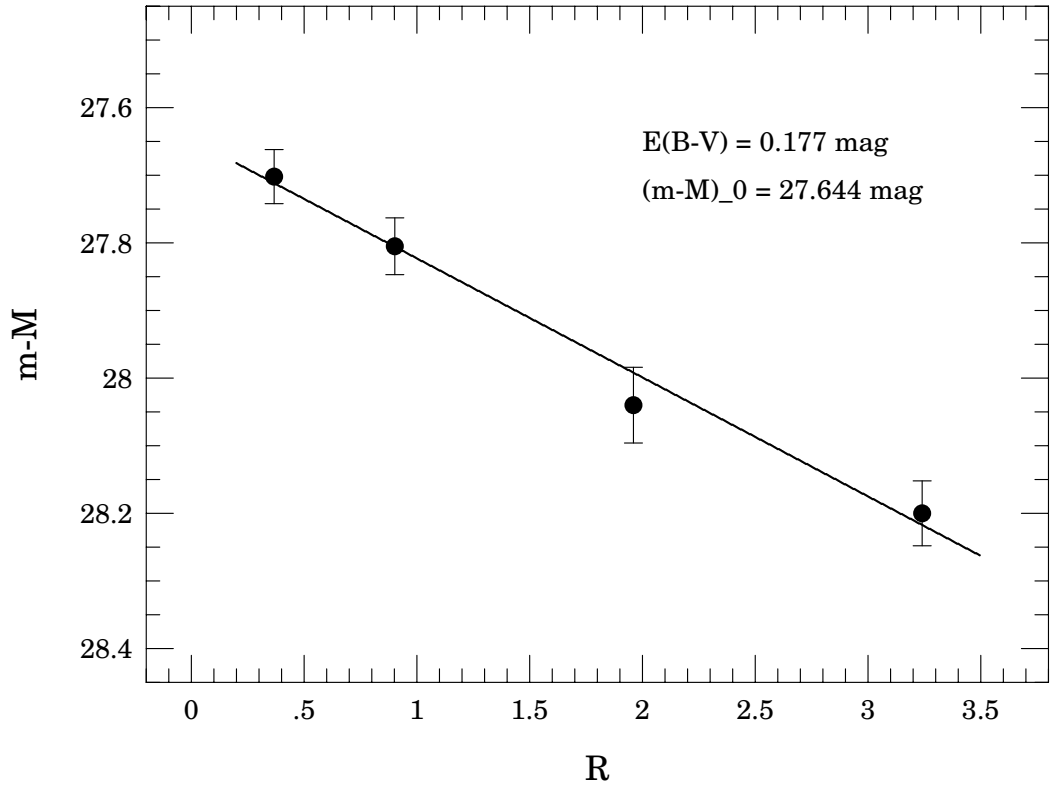}
\vspace{10cm}
\caption{Apparent distance moduli to NGC 247 derived in the VIJK photometric bands,
plotted against the ratio of total to selective extinction as adopted from
the Schlegel et al. reddening law. The intersection and
slope of the best-fitting line yield the true distance modulus and
reddening, respectively. The data in this diagram suggest that the Galactic reddening law
is a very good approximation for NGC 247 as well.}
\end{figure}

\clearpage
\begin{deluxetable}{c c c c c c c}
\tablewidth{0pc}
\tablecaption{Journal of the Individual J and K band Observations of NGC 247
Cepheids}
\tablehead{ \colhead{ID} & \colhead{J HJD} & \colhead{J}  & \colhead{$\sigma_{\rm J}$} 
& \colhead{K HJD} & \colhead{K} & \colhead{$\sigma_{\rm K}$} }
\startdata
cep006 & 3592.86819 & 21.216 & 0.068 & 3592.78328 & 20.753 & 0.098 \\
cep006 & 3661.62511 & 20.623 & 0.090 & 3661.60091 & 20.140 & 0.054 \\
cep008 & 3625.72588 & 20.771 & 0.054 & 3625.67282 & 20.288 & 0.075 \\
cep009 & 3657.69064 & 20.684 & 0.052 & 3657.60342 & 20.189 & 0.067 \\
cep012 & 3592.86819 & 20.433 & 0.033 & 3592.78328 & 19.969 & 0.044 \\
cep012 & 3661.62511 & 20.626 & 0.061 & 3661.60091 & 20.058 & 0.055 \\
cep015 & 3607.73856 & 20.846 & 0.036 & 3607.77029 & 20.454 & 0.058 \\
cep015 & 3693.69953 & 20.532 & 0.113 & 3693.65978 & 20.264 & 0.071 \\
cep016 & 3625.72588 & 20.671 & 0.048 & 3625.67282 & 19.839 & 0.062 \\
cep018 & 3625.72588 & 20.137 & 0.031 & 3625.67282 & 19.420 & 0.033 \\
cep019 & 3592.86819 & 20.072 & 0.038 & 3592.78328 & 19.361 & 0.062 \\
cep019 & 3661.62511 & 20.011 & 0.058 & 3661.60091 & 19.400 & 0.039 \\
cep021 & 3607.73856 & 19.861 & 0.035 & 3607.77029 & 19.164 & 0.040 \\
cep021 & 3693.69953 & 19.656 & 0.067 & 3693.65978 & 19.323 & 0.035 \\
cep023 & 3592.86819 & 18.873 & 0.020 & 3592.78328 & 18.356 & 0.034 \\
cep023 & 3661.62511 & 18.969 & 0.023 & 3661.60091 & 18.261 & 0.028 \\
\enddata
\end{deluxetable}

\clearpage

\begin{deluxetable}{c c c c c c}
\tablecaption{Intensity mean J and K magnitudes for 10 Cepheid variables in NGC 247}
\tablehead{
\colhead{ID} & \colhead{log P} &
\colhead{$<J>$} & \colhead{$\sigma_{\rm J}$} & \colhead{$<K>$} &
\colhead{$\sigma_{\rm K}$}\\ 
& \colhead{days} & \colhead{mag} & \colhead{mag} & \colhead{mag} &
\colhead{mag}
}
\startdata
cep006 &   29.585 &  20.908 &    0.064 &   20.326 &    0.063 \\ 
cep008 &   30.978 &  20.887 &    0.062 &   20.394 &    0.081 \\ 
cep009 &   32.114 &  20.781 &    0.060 &   20.314 &    0.073 \\ 
cep012 &   35.809 &  20.662 &    0.046 &   20.157 &    0.046 \\ 
cep015 &   41.393 &  20.569 &    0.066 &   20.160 &    0.055 \\ 
cep016 &   44.481 &  20.413 &    0.057 &   19.605 &    0.069 \\ 
cep018 &   63.505 &  20.194 &    0.043 &   19.487 &    0.045 \\ 
cep019 &   64.889 &  20.091 &    0.046 &   19.428 &    0.047 \\ 
cep021 &   69.969 &  19.754 &    0.048 &   19.213 &    0.040 \\ 
cep023 &  131.259 &  18.914 &    0.034 &   18.297 &    0.037 \\ 
\enddata
\end{deluxetable}

\begin{deluxetable}{cccccc}
\tablewidth{0pc}
\tablecaption{Reddened and Absorption-Corrected Distance Moduli for NGC
247 in Optical and Near-Infrared Bands}
\tablehead{ \colhead{} & $V$ & $I$ & $J$ & $K$ & $E(B-V)$ }
\startdata
 $m-M$                &   28.200 &  28.040 &  27.799 &  27.702 &   --  \nl
 ${\rm R}_{\lambda}$  &   3.24   &  1.96   &  0.902  &  0.367  &   --  \nl
$(m-M)_{0}$           &   27.625 &  27.693 &  27.639 &  27.637 &  0.177 \nl
\enddata
\end{deluxetable}

\end{document}